\documentstyle[aps]{revtex}
\begin{document}
\bibliographystyle{prsty}
\baselineskip=5.5mm
\parindent=7mm
\begin{center}
{\large {\bf \sc {Dynamical symmetry breaking, confinement with flat-bottom
 potential }}} \\[2mm]
Zhi-Gang Wang $^{1}$ \footnote{E-mail,wangzg@mail.ihep.ac.cn}, Shao-Long Wan  $^{2}$ \\
 $^{1}$Institute of High Energy Physics, P.O.Box 918 ,Beijing 100039,P. R. China \\
$^{2}$Department of Astronomy and Applied Physics, University of Science and \\
Technology of China, Hefei 230026

\end{center}

\begin{abstract}
 In this article, we calculate the dressed quark propagator with the
 flat bottom potential in the framework of the rain-bow
 Schwinger-Dyson equation. Then based on the nonperturbative
 dressed quark propagator, we calculate the $\pi$  decay constant
   and  the quark condensate. The $\pi$ decay constant is an important parameter
 in describing the interplay between dynamical symmetry breaking and
 confinement, while the quark condensate is an order parameter
 for dynamical chiral symmetry  breaking. To implement confinement,
 we prove that the dressed quark propagator has no poles on the real
 timelike $p^2$ axial, the absence of Kallen-Lehmann spectral
  representation  obviously precludes the existence of free quarks.

\end{abstract}



\section{Introduction}

The standard model gives highly successful descriptions
of hadronic physics in terms of the basic force.
Yet its key parameters, the masses of the quarks and leptons
 are specified by the coupling of the Higgs bosons, which is
 undetermined. It is possible that the dynamics of the
fundamental gauge theories themselves generate the masses of
all the matter fields, or in other word, dynamical chiral symmetry breaking
(for quarks),
which corresponding to the absence of the low mass scalar partners
of the $\pi$ \cite{Miransky}. Quantum chromodynamics (QCD) is the
appropriate theory for describing the strong interaction at high energy.
However, the lack of observation of free quarks and gluons leads to the
assumption of colour confinement that only quantities which transform
 as colour singlets can be physical observable.
One may expect that the nonperturbative infrared behaviour of gluon
propagator which may lead to a linear or oscillator potential implements
confinement, however, till now, all theoretical studies based on QCD can not give
satisfactory explanation. Hence we take an equivalent statement that
the propagator of a colored state should have no singularities on
the real time-like $p^2$ axial, the absence of Kallen-Lehmann spectral
representation precludes the existence of free quarks \cite{Roberts1,Bjorken}.

In fact, dynamical symmetry breaking and confinement are two
crucial features of QCD, however, they correspond to two very
different energy scales. The possible interplay of the two dynamics
lies in the following two facts, the first one is that
on large energy scale the chiral massive quark interacts
with antiquark via a confining potential and creates a new
quark mass, constituent quark mass; while the second one can be
attributed to the double role of the $\pi$, as both Nambu-Goldstone
boson and  $q\bar{q}$ bound state.

 The Schwinger-Dyson equation, in effect the functional
 Euler-Lagrange equation of quantum field theory, provides a natural
  framework for investigating the nonperturbative properties  of
  quark and gluon Green's functions. By studying the evolution
  behaviour and analytic structure of the dressed quark propagator,
  one can obtain valuable information about the dynamical symmetry
  breaking phenomenon and confinement.

Although lattice calculations are rigorous in view of QCD,
they suffer from lattice artifacts and uncertainties, such as Gribov
copies, boundary conditions and so on. Furthermore,  current technique
can not give reliable result below $1 GeV$, where the most interesting
and novel behaviour is expected to lie.

 It would seem therefore that at the present time Schwinger-Dyson (SD)
  equation  is the most reliable tool for studying the infrared
  behaviour of the quarks and gluons propagators in the continuum limit,
  while it has its own shortcomings.

The flat-bottom potential (FBP) is a sum of Yukawa potentials, which
  not only  satisfies gauge invariance, chiral invariance and fully
relativistic covariance, but also suppresses the singular point which the
 Yukawa potential has. It works well in understanding the meson
 structure, such as electromagnetic form factor, radius, decay constant
 \cite{Wang}.

Therefore, it is interesting to examine dynamical chiral symmetry
breaking and quark confinement with the FBP.

The article is arranged as follows: in section 1, introduction;
in section 2, infrared behaviour of the gluon  propagator;
in section 3, dynamical chiral symmetry breaking;
in section 4, quark confinement; in section 5, flat bottom
potential and Schwinger-Dyson equation; in section 6,
conclusion and discussion.

 \section{Infrared behaviour of gluon propagator }
The infrared structure of the gluon propagator has important
implication for quark confinement. One might expect that the behaviour
of the quark interaction in region of small  spacelike $p^2$
determines the long range properties of the $q\bar{q}$ potential and hence
implements  confinement.

From the discussion of previous section, we can see that the SD
equation is the ideal tool for studying the infrared behaviour of
the gluon propagator. Here we make a brief skeleton of the results
based on Landau gauge and axial gauge studies.

It is impossible to solve an infinity series of coupled SD equations
for gluons and quarks (ghosts), so the authors have to make truncations
both in Landau gauge and axial gauge. In those studies, the quark loop
contributions are neglected and higher order vertexes  are taken
as bare or dressed based on the analysis of corresponding
Ward-Takahashi Identity.

In axial gauge studies, one makes further assumption that the
dressed gluon propagator has the same tensor structure as the free
one. That is in the following equations let
$F_{2}=0$ \cite{Baker},
\begin{eqnarray}
D_{\mu \nu}(p,\gamma)& = & F_{1}(p,\gamma)M_{\mu \nu}(p,n)+F_{2}(p,\gamma) N_{\mu \nu}(p,n),  \nonumber \\
M_{\mu \nu}(p,n)&=&\delta_{\mu \nu}-\frac{p_{\mu}n_{\nu}+p_{\nu}n_{\mu}}{p \cdot
n}+n^2 \frac{p_{\mu}p_{\nu}}{(p \cdot n)^2}, \nonumber \\
N_{\mu \nu}(p,n) &=& \delta_{\mu \nu}- \frac{n_{\mu}n_{\nu}}{n^2},
\end{eqnarray}
here $\gamma$ is gauge parameter and n is an arbitrary four vector.
After solving gluon SD equation one get the conclusion that the
gluon propagator may have a double pole at the origin, implying  an area law behaviour
 of the Wilson loop, which is often regarded as a signal of confinement \cite{West1}.
 Further more , the Fourier transform of the time-time component of
 the gluon propagator is related to the one-gluon exchange contribution to
 the potential between static colour charge and a $\frac{1}{p^4}$ infrared
 behaviour does generate a linearly rising potential at large distance \cite{Braun1}.
 However,  a general studies of the
 properties of the spectral representation of the gluon propagator
 in axial gauge show that the coefficient of the $\delta_{\mu \nu}$ term
 cannot be more singular  then $\frac{1}{q^2}$ \cite{West2}.
 This suggests that neglecting the $F_{2}$ term is a poor approximation
 because there  maybe  cancellation in the infrared between $F_{1}$ and $F_{2}$.

  In Landau gauge studies \cite{Braun1,Mandelstam}.
  Suppose neglecting ghost loop  will not  remove  the dominant
  infrared contributions in QCD also leads to a  $\frac{1}{p^4}$
  behaviour for the gluon propagator in the origin.
 Other studies  suggest that fermion loop may suppress the infrared
 singularity in the gluon propagator, the double pole found
 in the pure-gauge sector may be removed by the nonperturbative fermion
  loop corrections \cite{Braun3,Alkofer,Roberts}. Recently, the gluon
  SD equation   studies including ghost loop result in a  infrared vanished
   gluon propagator  and a strong infrared enhanced ghost propagator
  \cite{Smekal,Atkinson} while lattice calculations
support this type conclusion \cite{Leinweber}.
In fact, the truncations applied in these studies may impair their
persuasion. Furthermore, in some studies of this type, unphysical
particle-like singularities occur in the  dressed
ghost-gluon and 3-gluon vertices  and
the infrared vanished gluon propagator can neither
produce  dynamical chiral symmetry breaking
nor compliment  confinement \cite{Smekal,Hawes}.

We can take a simple example  by analyzing the renormalization group
equation for the coupling constant. There exists a competing mechanism
 between the self-interaction of gauge boson and the interaction
 of  boson-fermion, which accounts for the different evolution
 behaviours of the coupling constants with momentum for QCD and
 QED \cite{Marciano}. So neglecting one another may lead to
 very different results.

All theoretical studies based on QCD are suffered from shortcomings in
one or another way \cite{Alkofer}, so they can not give satisfactory
infrared behaviour for gluon propagator which may lead to a linear or
 oscillator potential implements confinement.

Here we take an alternate description of confinement based on the
analytic structure of the dressed quark propagator,
if the quark Green's function is free of singularities on the
real timelike $p^2$ axial , then denies the existence of a single
particle spectrum and precludes free quarks. Although by no means a
sufficient condition for the formulation of finite size $q \bar{q}$
states, it is an essential ingredient in describing of confinement
based on a bold assumption that interaction between dressed quarks
which does not rise to infinity  at large distance.
The interaction between dressed quarks may diminish
at separation beyond the characteristic length scale at which
point the quark Green's function approaches its vacuum values,
but the vacuum's repulsive force due to the absence of a mass pole in
the dressed quark propagator again compresses the quarks
together.

So the traditional model of confinement based on an effective rising
potential can be interpreted as the result of  combination of the
repulsive interaction between  quark-vacuum and the attraction of
dressed quark-quark with increasing separation. Although there maybe
a long way to go before the repulsive quark-vacuum interaction can be
quantified, this description has proven successful in the formulation
of a confining  nontopological soliton model for baryons \cite{Frank1}. In this article,
the quark propagator calculated with the FBP does free singularities
on the real timelike $p^2$ axial and implements quark confinement.

\section{Dynamical Chiral Symmetry Breaking}
 The ultraviolet stability of QCD at large spacelike momentum
 makes a straightforward definition of the chiral limit possible,
 in a manifest contrast to strong coupling QED, whose rigorous
 definition remains an instructive challenge.   The $\pi$ decay
 constant $f_{\pi}$ is an important parameter in describing
  the interplay between dynamical symmetry breaking and confinement,
  while the quark condensate serves as an order
  parameter for dynamical chiral symmetry breaking.
  In the chiral limit, the $\pi$ plays a double role both as
  Nambu-Goldstone boson and $q \bar{q}$ bound state, which lead to the
  result  that the scalar part of the inverse dressed quark propagator coincides with
  the approximated $\pi$'s Bethe-Salpeter function. In the following,
   we calculate the $\pi$ decay constant , the quark condensate and
    examine the dynamical chiral symmetry breaking phenomenon.

 \subsection{$\pi$ Decay Constant}
Neglecting ghost, the Ward-Takahashi Identity (WTI) for  quark-gluon
vertex in the presence explicit quark masses can be written as \cite{Itzykson}:
\begin{eqnarray}
k_{\mu} \Gamma^{\mu}_{5}(p',p)=
\frac{\tau}{2}[S^{-1}(p')\gamma_{5}+\gamma_{5}S^{-1}(p)]+
2im \Gamma_{5}(p',p) .
\end{eqnarray}

 In the limit of exact chiral symmetry, the WTI reduces to
 \begin{eqnarray}
 k_{\mu} \Gamma^{\mu}_{5}(p',p)
  \stackrel {k \rightarrow 0} {\longrightarrow }-\tau
 \gamma_{5}B(p^2).
 \end{eqnarray}
 Note that the usual perturbative isovector axial vector vertex is $\frac{\tau}{2}\gamma_{\mu}\gamma_{5}$.

 In the chiral limit and take $k \rightarrow 0$, the axial vector quark
 vertex becomes completely dominated by the pseudoscalar
 coupling of a massless $\pi$ to the quark and the subsequent weak
 decay of the $\pi$ into an axial vector current.
 In fact, the mass shell residue of the axial vector quark vertex
  ($\Gamma^{5}_{\mu}$) is the $\pi$  Bethe-Salpeter $(\Gamma_{\pi})$ amplitude,
  in the chiral limit can provide a direct connection to the quark
  propagator \cite{Jackiw}. The function
  $\Gamma^{5}_{\mu}(p+\frac{k}{2},p-\frac{k}{2})\ \ (\Gamma^{5})$
  contains many Dirac structures, such as $\gamma_{5} \gamma_{\mu}$
  $(\gamma_{5})$, $\gamma_{5}  \gamma_{\mu} \gamma \cdot k$
  $ (\gamma_{5}\gamma \cdot k)$,
  $\gamma_{5} \gamma \cdot p p_{\mu}$
  $ (\gamma_{5} \gamma \cdot p p \cdot k)$, $\cdots$ and
  regular term, singular term.
  Exclude the singular term $(\Gamma_{\pi})$ ,
  the other terms are  supposed
  to give small contributions and neglected.
  If we solve the Schwinger-Dyson equation  for
  the vertex, we can see their relative weights.  In the chiral limit,
  the $\gamma_{5}$ component of $\Gamma_{\pi}$ makes a good approximation,
  though it has many Dirac structures, such as $\gamma_{5}$,
  $\gamma_{5}\gamma \cdot k$ and so on.   The $\pi$ decay constant
  $f_{\pi}$ is defined by  the axial transition amplitude for the
  on-shell $\pi$,

\begin{equation}
\langle 0| A^{m \mu}_{5}(0)|\pi^{n}(k)
\rangle=if_{\pi}k^{\mu}\delta^{mn}.
\end{equation}
When $k \rightarrow 0$, we obtain
\begin{eqnarray}
i\Gamma^{\mu}_{5}(p',p)  \stackrel{k\rightarrow 0}
{\longrightarrow}  i\Gamma^{m}_{5}(p',p) \frac{i}{k^2} i \langle 0 |A^{\mu}_{5}|\pi^{m}(-k)
\rangle
 \stackrel{k \rightarrow 0}{\longrightarrow}
-\Gamma_{5}(p',p)f_{\pi} \frac{k^{\mu}}{k^2}.
\end{eqnarray}
Multiplying Eq.(5) with $k_{\mu}$ on both sides,
\begin{equation}
k_{\mu} \Gamma^{\mu}_{5}(p',p)=if_{\pi}\Gamma_{5}(p',p) .
\end{equation}
When combined with Eq.(3), we obtain the Goldberger-Treiman relation
for the quark-pseudoscalar vertex,
 \begin{equation}
 \Gamma_{5}(p,p)=i\tau \gamma_{5} \frac{B(p^2)}{f_{\pi}}
 =  i\tau \gamma_{5} \frac{A(p^2)m(p^2)}{f_{\pi}}.
 \end{equation}
The simplest generalization of Eq.(7) is
\begin{equation}
\Gamma_{5}(p',p)=i\tau \gamma_{5}
\frac{B(p'^2)+B(p^2)}{2f_{\pi}}.
\end{equation}

Based on above discussion, it is possible to obtain an expression for
$f^2_{\pi}$ from the integral equation for the $\pi$ decay constant.

\begin{eqnarray}
i\langle 0|A^{m}_{\mu}(0)|\pi^{n}(k) \rangle  =- \int \frac{d^4 p}{(2\pi)^4}tr \{ iS(p+k)i \Gamma^{m}_{5}(p+k,p)iS(p) i
\frac{\tau^{n}}{2} \gamma_{\mu} \gamma_{5} \}.
\end{eqnarray}

As usual, the factor $-1$ on the right hand side arises from the quark
loop. We need to take the axial-vector vertex as a perturbative one
in order to avoid double counting.
The expression for $f^{2}_{\pi}$ can be written as
 \begin{equation}
 f^{2}_{\pi}=-12i \int \frac{d^4 p}{(2 \pi)^4}
 \frac{B(p^2)} {(A(p^2)p^2-B^{2}(p^2))^2}
  \{A(p^2)B(p^2)+\frac{p^2}{2}(B\frac{d A(p^2)} {dp^2}-A\frac{d B(p^2)} {dp^2}) \}.
 \end{equation}

In the limit $A \equiv 1$, we can recover the Pagels and Stokar
formula  \cite{Cornwall}. Our numerical results based on the
 quark propagator obtained from the SD equation studies
 with the FBP do give satisfactory result $f_{\pi}=94MeV$, we can
thus fit the parameters in the FBP.

There is another expression for the $\pi$ decay constant which
determined from global colour model \cite{Frank0,Tandy}.
\begin{equation}
f^2_{\pi}=\frac{3}{8 \pi^2} \int_{0}^{\infty} ds s B(s)^2
 \{ \sigma^2_v-2[\sigma_{s}\sigma'_{s}+s \sigma_{v}\sigma'_{v}]
 -s[\sigma_{s} \sigma''_{s}-\sigma'^{2}_{s}]-s^{2}[\sigma_{v} \sigma''_{v}-\sigma'^{2}_{v}]\}
\end{equation}
where
\begin{eqnarray}
 \sigma_{v} = \frac{A(s)}{s A(s)^{2}+B(s)^{2}}, \ \ \  \sigma_{s} = \frac{B(s)}{s
 A(s)^{2}+B(s)^{2}}.
\end{eqnarray}
In our calculation. the value of $A(p)$ is far from 1,  we
prefer Eq.(10) rather than Eq.(11), as Eq.(11) is exact only in
the limit $A(p)\equiv 1$ \cite{Frank0,Tandy}.

\subsection{Quark Condensate}
 The quark propagator is defined as
\begin{equation}
S(x)=\langle 0 |T[q(x) \bar q(0)]|0 \rangle  .
\end{equation}
where q(x) is the quark field and T the time-ordering operator.
For the physical vacuum consists of both perturbative and
nonperturbative parts, so the quark propagator S(x) can be divided into a perturbative
and a nonperturbative part as the following:
\begin{eqnarray}
S(x)= S_{PT}(x)+S_{NP}(x) .
\end{eqnarray}
In the nonperturbative vacuum, the normal-ordered product $S_{NP}(x)$
does not vanish.
The widely used nonlocal quark condensate  $\langle0| : \bar q(x) q(0) :|0 \rangle$ is given
 by the scalar part of the Fourier transformed inverse quark
propagator,
\begin{eqnarray}
\langle0|:\bar{q}(x)q(0):|0\rangle_{\mu}&=&(-4N_{c})\int^{\mu}_{0}\frac{d^4 p}{(2\pi)^4}\frac{B(p^2)}
{p^2A^2(p^2)+B^2(p^2)}e^{ipx}.
\end{eqnarray}
In the limit $x \rightarrow 0$, we can
obtain the expression for the local quark condensate,
\begin{equation}
\langle 0|: \bar q(0) q(0) :|0 \rangle_{\mu} = -\frac{12}{16 \pi^2}
\int_{0}^{ \mu} d s s \frac {B(s)}{sA^2(s)+B^2(s)}.
\end{equation}
In Ref.\cite{Kisslinger2} the nonlocal condensate is put in the
following form :
\begin{equation}
\langle 0|: \bar q(x) q(0) :|0 \rangle=g(x^2)\langle0| : \bar q(0) q(0) :|0 \rangle
,
\end{equation}
where $g(x^2)$ is the vacuum non-locality of the nonlocal quark
condensate.
The nonlocal properties of the vacuum condensates are
 of principal importance in the study of the distributions of quarks
 and gluons in hadrons \cite{Mikhailov1}. Physically, it means that the vacuum condensates of
 quark and gluon can flow through the vacuum with nonzero
 momentum.

The effects from hard gluonic radiative corrections to the
 quark propagator are connected to a possible change of the renormalization
 scale $\mu$ at which the condensates are defined. Those effects are of minor
 importance for the nonperturbative effects in the low and medium energy regions,
 and neglected in our studies . In this article, we take the rain-bow SD equation,
 so it is not a renormalizable interaction. As isolated closed quark loop integral,
 the quark condensate has only one energy
scale, the  momentum cut off, $\mu$, at which the quark condensate
is defined. In our studies, the energy scale $\mu$ is implicitly determined
by the effective gluon propagator, we can use Eq.(16) as the definition
for the quark condensate. However, presently, the nonperturbative technique can not prove
the relation
 \begin{equation}
\int_{0}^{ \mu} d s s \frac {B(s)}{sA^2(s)+B^2(s)}
\sim (Log(\mu^2/\Lambda^2_{QCD}))^d,
 \end{equation}
at low energy scale. Here $d$ is the anomalous dimension and takes
the value $d=\frac{12}{33-2n_{f}}$.

If we take the limit $\mu \rightarrow \infty$ and
  assume ultraviolet
 dominating, the above relation $\langle \bar{q} q \rangle_{\mu} \sim
 (Log(\mu^2/\Lambda^2_{QCD}))^d$ is indeed the case, although the
 quark condensate is only defined at low energy scale.

 Operator product expansion has proven that at large Euclidean momentum,
 the effective quark mass evolves  as
 \begin{equation}
 m(-Q^2)_{Q^2 \rightarrow \infty} = \frac{c}{Q^2} \{ Log(Q^2 / \Lambda^2_{QCD})
 \}^{d-1} +m(\mu) \{
 \frac{Log(\mu^2 / \Lambda^2_{QCD})}{Log(Q^2 /
 \Lambda^2_{QCD})}\}^{d},
 \end{equation}
here $c=-\frac{4 \pi d}{3} \frac{\langle 0|:\bar{q}q
:|0\rangle}{[Log(\mu^2/\Lambda^2_{QCD})]^d}$ is some constant
independent of $\mu$ \cite{Politzer,Gasser}, this implies that
$\langle0|: \bar{q} q:|0 \rangle_{\mu} \sim
[Log(\mu^2/\Lambda^2_{QCD})]^d$, but not the definition of Eq.(16) .

On the other hand, the evolution of quark condensate with $\mu$ can be
determined from  Gell-Mann-Oakes-Renner relation and renormalization group equation,
 \begin{equation}
 \langle 0|:\bar{q}q :|0\rangle_{\mu} =\{ \frac{Log(\mu^2/\Lambda^2_{QCD})}{Log(\Lambda^2/\Lambda^2_{QCD})}
 \}^{d}\langle 0|:\bar{q}q :|0\rangle_{\Lambda}.
 \end{equation}
 If we take into
 account the hard radiative corrections to the quark propagator ,
 there is a more nature definition for the quark
 condensate (in Euclidean space time) \cite{Maris}:
 \begin{equation}
 \langle 0|:\bar{q}q :|0\rangle_{\mu} =\{ \frac{Log(\mu^2/\Lambda^2_{QCD})}{Log(\Lambda^2/\Lambda^2_{QCD})}
 \}^{d} \frac{12}{16 \pi^4} \int_{0}^{\Lambda} d^4 p \frac{B(p^2)}{A(p^2)^2
 p^2+B(p^2)^2},
 \end{equation}
  In the second definition, we can take
 the cut off be $\Lambda \rightarrow \infty$, however, at large momentum,
 the quark propagator approximates its asymptotic free form which should be subtracted.
 From Fig.1, we can see that the cut off $\Lambda =2GeV$  makes a sound approximation.

 In this article, $\mu$ is taken to be 1 $GeV$.
If we ignore the hard radiative corrections to the quark propagator,
from Eq.(16), the numerical result gives the quark condensate $ \langle 0| \bar{q}q |0
\rangle^{\frac{1}{3}}=-218MeV$.
If we take into account the hard radiative corrections, from Eq.(21), we obtain:
 \begin{eqnarray}
\Lambda_{QCD}=150MeV, \ \ \  \langle  \bar q(0) q(0)
\rangle^{\frac{1}{3}}=-255MeV; \\ \nonumber
 \Lambda_{QCD}=200MeV,\ \ \ \langle \bar q(0) q(0)
 \rangle^{\frac{1}{3}}=-253MeV; \\ \nonumber
 \Lambda_{QCD}=250MeV, \ \ \ \langle  \bar q(0) q(0)
 \rangle^{\frac{1}{3}}=-251MeV \nonumber.
 \end{eqnarray}
All our numerical results  are compatible with the results of the
QCD sum rule approach, and large
 enough to give dynamical chiral symmetry  breaking. The
 definition of Eq.(21) is better.

\section{quark Confinement}
To demonstrate that quark confinement arises from the absence of
singularities on the real timelike $p^2$ axial, let us first make an
observation on the large Euclidean time behaviour of the free
fermion propagator \cite{Roberts2}.
\begin{eqnarray}
S(x) & & = \int \frac{d^4 k}{(2 \pi)^4} e^{i k \cdot x} \frac{m-i \gamma \cdot
k}{k^2+m^2},
\end{eqnarray}

The spatially averaged Schwinger function is a particularly
insightful tool. Consider the fermion propagator and let $T=ix_{4}$ represent
Euclidean time, then
\begin{eqnarray}
\sigma_{s}(T)  = \int d^3 x \frac{1}{4} tr_{D} S(x,T)= \frac{1}{2}e^{-mT}.
\end{eqnarray}
Hence the free fermion's mass can easily be obtained from the
large T behaviour of the spatial average:
\begin{equation}
\lim_{T \rightarrow \infty} \log \sigma_{s}(T)=-mT.
\end{equation}
This is just the approach used to determine the  bound state masses in
simulations of Lattice QCD.

In order to demonstrate that the confinement of quarks, we have to
study the mass function of the quark and prove
 that there no poles on the real timelike like $p^2$ axial.
So it is necessary to perform an analytic continuation of
the dressed quark propagator from Euclidean space into Minkowski
space $p_{4} \rightarrow ip_{0}$.
However, we have no knowledge of the singularity structure of
quark propagator in the whole complex plane. One can take an
alternative safe procedure and stay completely in Euclidean
space avoiding analytic continuations of the dressed
propagators \cite{Burkardt}. Again we  take the Fourier transform
 with respect to the Euclidean time T. It is sufficient to consider
 the scalar part $S_{s}$,
 \begin{eqnarray}
 S^{*}_{s}(T) & & =\int_{-\infty}^{+ \infty} \frac{dq_{4}}{2 \pi}
 e^{iq_{4}T}S_{s}, \nonumber \\
 & & =  \int_{-\infty}^{+ \infty} \frac{dq_{4}}{2 \pi} e^{iq_{4}T}
 \frac{B(q^2)}{q^2A^2(q^2)+B^{2}(p^2)}.
 \end{eqnarray}
 If S(p) had a pole at $p^2=-m^2$, the Fourier transformed
  $S^{*}_{s}(T)$ would fall off as $e^{-mT}$ for large T or
  $\log{S^{*}_{s}}=-mT$.

In our calculation, for large $T$, the values of $S^{*}_{s}$ is negative,
except occasionally a very very small fraction positive values which
can be safely neglected. We can express
$S^{*}_{s}$ as $|S^{*}_{s}|e^{i n\pi}$, $n$ is an odd integer.
$\log{S^{*}_{s}}=\log|{S^{*}_{s}}|+in \pi=-mT$.
If we neglect the imaginary part, we find that when the Euclidean time T is
large, there indeed exists a crudely approximated linear function with
respect to T, which is shown in Fig.2. However, such a behaviour is hard
to acquire  physical explanation. Here the word 'crudely' should be
understand in the linearly fitted sense, to be exact, there is not
linear function. This precludes the existence  of free quarks.

\section{Flat Bottom Potential and Schwinger-Dyson Equation}
The FBP is a sum of Yukawa potentials which is an analogous to the
exchange of a series of particles and ghosts with different
masses (Euclidean Form),
\begin{equation}
G(k^{2})=\sum_{j=0}^{n}
 \frac{a_{j}}{k^{2}+(N+j \rho)^{2}}  ,
\end{equation}
where $N$ stands for the minimum value of the mass, $\rho$ is their mass
difference, and $a_{j}$ is their relative coupling constant.
In Fig.3, we can see the curve of $G(p^2)$ is a gauss-like
distribution. Due to the particular condition we take for the FBP,
there is no divergence in solving the SD equation.
In its three dimensional form, the FBP takes the following form:
\begin{equation}
V(r)=-\sum_{j=0}^{n}a_{j}\frac{{\rm e}^{-(N+j \rho)r}}{r}  .
\end{equation}
In order to suppress the singular point at $r=0$, we take the
following conditions:
\begin{eqnarray}
V(0)=constant, \nonumber \\
\frac{dV(0)}{dr}=\frac{d^{2}V(0)}{dr^{2}}=\cdot \cdot
\cdot=\frac{d^{n}V(0)} {dr^{n}}=0    .
\end{eqnarray}
So we can determine $a_{j}$ by solve the following
equations, inferred from the flat bottom condition Eq.(29),
\begin{eqnarray}
\sum_{j=0}^{n}a_{j}=0,\nonumber \\
\sum_{j=0}^{n}a_{j}(N+j \rho)=V(0),\nonumber \\
\sum_{j=0}^{n}a_{j}(N+j \rho)^{2}=0,\nonumber \\
\cdots \nonumber \\
\sum_{j=0}^{n}a_{j}(N+j \rho)^{n}=0 .
\end{eqnarray}
As in  previous literature \cite{Wang}, $n$ is set to b 9.

In the rainbow approximation, the SD equation takes the following form:
\begin{equation}
S^{-1}(p)=\gamma \cdot p - \hat{m}+\frac{16 \pi i}{3}\int
\frac {d^{4}k}{(2 \pi)^{4}} \gamma_{\mu}
S(k)\gamma_{\nu}G^{\mu \nu}(k-p),
\end{equation}
where
\begin{eqnarray}
S^{-1}(p)&=& A(p^2)\gamma \cdot p-B(p^2)\equiv A(p^2)
[\gamma \cdot p-m(p^2)], \\
G^{\mu \nu }(k)&=&-(g^{\mu \nu}-\frac{k^{\mu}k^{\nu}}{k^2})G(k^2),
\end{eqnarray}
and $\hat{m}$ stands for an explicit quark mass-breaking term.
With the explicit small mass term, we can preclude the zero
solution for $B(p)$ and in fact there indeed exist a bare current
quark mass. Here we take Landau gauge.
This dressing comprises the notation of constituent quarks by
 providing a mass $m(p^2)=B(p^2)/A(p^2)$, which is corresponding to
 dynamical symmetry breaking. Because the form of
 the gluon propagator $g^2G(p)$ in the infrared region is unknown
 (as we have discussed in section 2),
 one often uses model forms as input in the previous studies
  of the rainbow SD equation
  \cite{Tandy,Roberts}.

In this article, we assume that a Wick rotation to Euclidean variables is
allowed, and perform a rotation analytically continuing $p$ and $k$
into the Euclidean region where them can be denoted by $\bar{p}$ and $\bar{k}$,
 respectively. Alternatively, one can derive the SD equation from the
 Euclidean path-integral formulation of the theory, thus avoiding
 possible difficulties in performing the Wick
 rotation $\cite{Stainsby}$ . As far as only numerical results are concerned,
  the two procedures are equal.

The Euclidean SD equation can be projected into two coupled integral
equations for $A(\bar{p}^2)$ and $B(\bar{p}^2)$. For simplicity, here
ignore the bar on $p$ and $k$.
 Numerical values for $A(p^2)$, $B(p^2)$ and $m(p^2)$
are shown in Fig.[1] .\\

\section{Conclusion and Discussion}

In this article, we calculate the dressed quark propagator with the
 FBP in the framework of the rain-bow SD equation. The dressed quark
 propagator exhibits a dynamical symmetry  breaking phenomenon
 and gives a constituent quark mass about 490 MeV ($m(m^2)$), which is
 larger then  the value of commonly used constituent quark mass 350 MeV in
 the chiral quark model. Then based on dressed quark propagator, we
 obtain the quark condensate $\langle 0|\bar{q}q|0\rangle$, the $\pi$ decay constant.
The calculated $f_{\pi}$ and $\langle 0|\bar{q}q|0\rangle$ are
compatible with experimental and theoretical works respectively.
Though the value of $f_{\pi}$, we can fit the parameters:
 $N=1.0\times 200MeV $, $V(0)=-27.0\times 200MeV$,
 $\rho=5.0\times 200MeV$, $m_{u}=m_{d}=8 MeV$ and the large momentum
cut-off $L=630\times 200MeV$ in solving the SD equation.

The $\pi$ decay constant $f_{\pi}$ is an important parameter in describing
the interplay between dynamical symmetry breaking and confinement,
while the quark condensate serves as an order
parameter for dynamical chiral symmetry breaking.
In the chiral limit, the $\pi$ plays a double role both as
Nambu-Goldstone boson and $q \bar{q}$ bound state, which lead to the result
that the scalar part of the inverse dressed quark propagator coincides with
the approximated $\pi$'s Bethe-Salpeter function.

Till now, all theoretical studies based on QCD are suffered from
shortcomings in one or another way \cite{Alkofer}, so they can not
give satisfactory infrared behaviour for gluon propagator which
may lead to a linear or  oscillator potential implements confinement.
In this article, we take an alternative description by studying the
analytic structure of the quark propagator. If the dressed quark
propagator is free singularities on the real timelike $p^2$ axial,
although the attraction between quarks may diminish at
a characteristic distance, but the vacuum repulsive force due to the
absence of mass poles compresses the quarks together.
The traditional confinement based on the arising potential between
quarks as the separation becomes long can be realized as
combination of both the attractive and repulsive forces.
In our calculation, we take a Fourier transform with respect
Euclidean time T, which is often used in lattice calculation
to extrapolate physical masses, and find that at large time T,
there does not exist a physical quark mass, which may
preclude the existence of physical free quark spectrum.

In calculation, the coupled integral equations for the quark
propagator functions $A(p^{2})$ and $B(p^{2})$
are solved numerically by simultaneous iterations. The iterations converge
rapidly to a unique stable solution of propagator functions and independent
the initial guesses. The propagator functions $A(p^{2})$ and
$B(p^{2})$ are shown in Fig.[1] , at small $p^{2}$ , $A(p^{2})$ differs from
the value 1 appreciably, while it tends to 1 for large $p^{2}$.
We find that at small $p^{2}$ , $m(p^{2})$ is greatly re-normalized,
 while at large $p^{2}$,
it takes asymptotic behaviour. For $u$ and $d$ quark, $m(0)=632 MeV$,
 the connection of $m(p)$ to constituent masses is somewhat less direct
and is precise only for heavy quarks. For heavy quarks,
$ m_{constituent}(p)=m(p=2m_{constituent}(p))$ , for light quarks ,
it only makes a crude estimation \cite{Politzer}. At about $p=1GeV$,
the mass function grows rapidly as the momentum decreases, that is
an indication of dynamical symmetry breaking.  There is a popularly  used
constituent quark definition, $m(p^2)^2=p^2 $ \cite{Tandy}. Here $p^2$ is
Euclidean momentum square and $m(m^2)=490 MeV$. Our result is
larger than the usually used constituent quark mass $m=350 MeV$,
however, it gives a good description of dynamical symmetry breaking.

The  phenomenological FBP can be extended  to other system, such as $q$$\bar{q}$ system, $\bar{q}$$Q$ system, vector meson and $qqq$ system or by applying the results to the
calculation of quantities and processes requiring detailed
knowledge of the quark propagator .

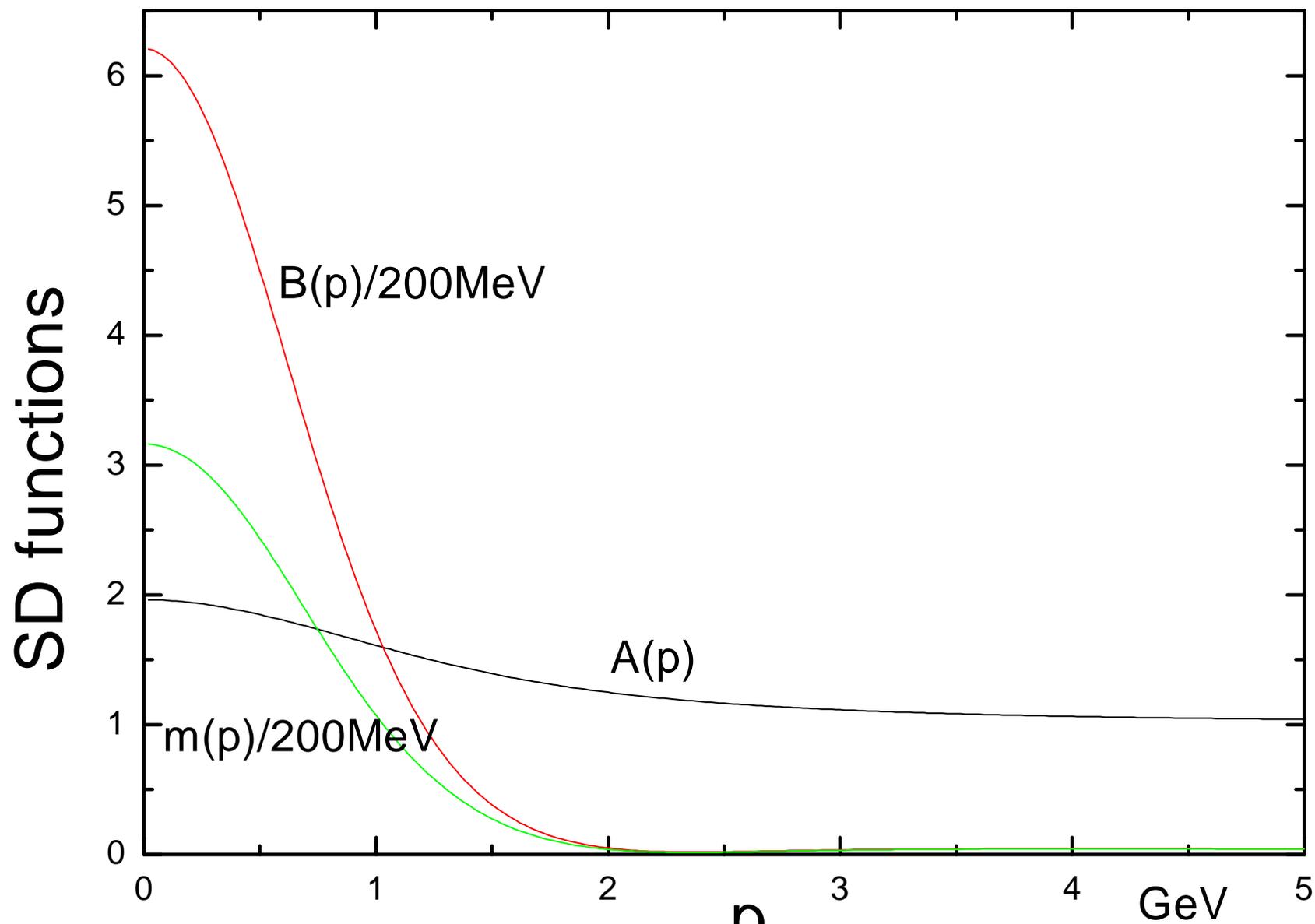
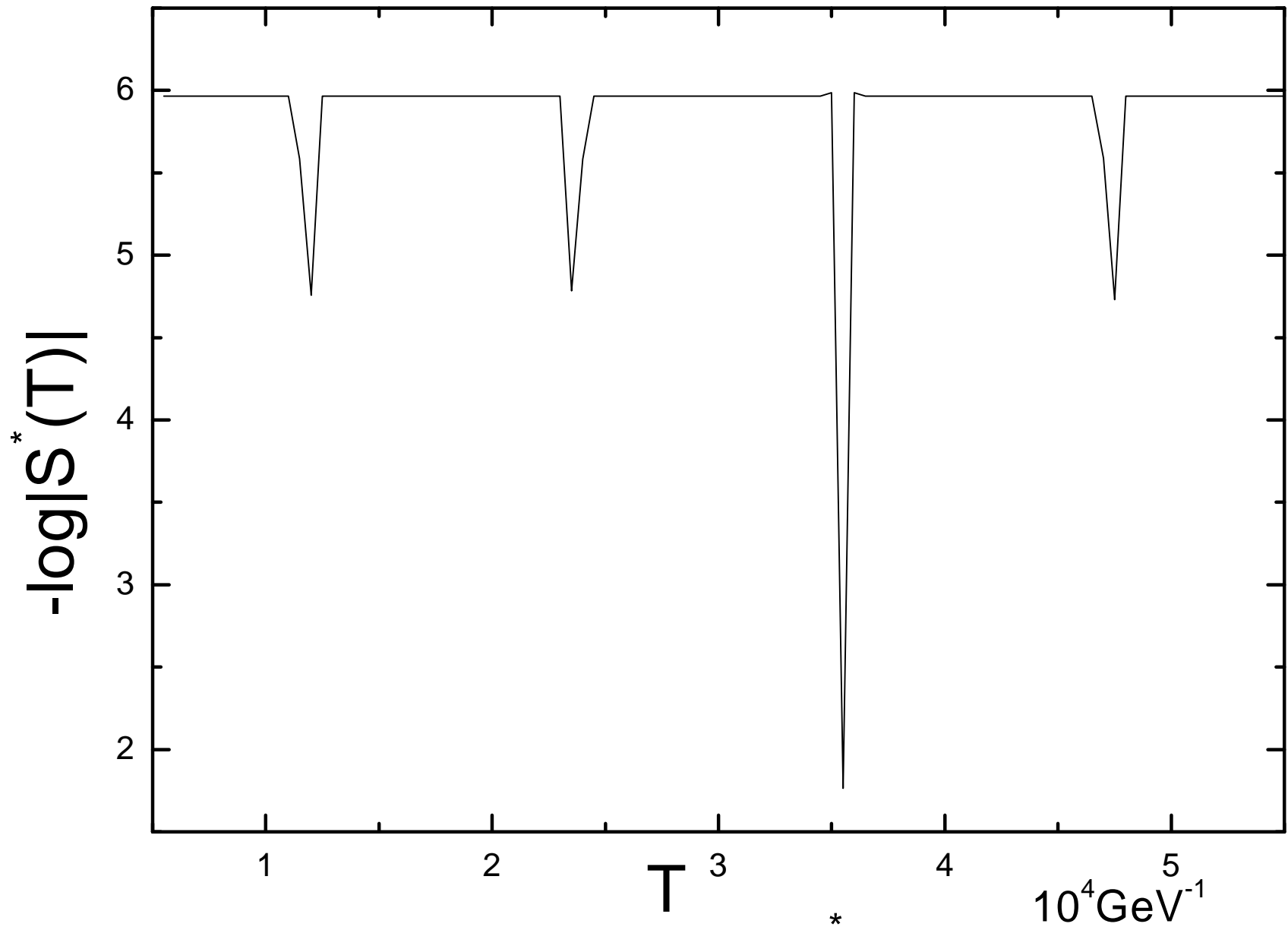
\end{document}